\documentclass[review]{elsarticle}
\usepackage{lineno,hyperref}
\usepackage{comment}
\usepackage[utf8]{inputenc}
\usepackage[T1]{fontenc}
\usepackage{mathptmx}
\usepackage{verbatim}
\usepackage{physics}
\usepackage{fancyhdr}
\usepackage{calc}
\usepackage{geometry}
\usepackage{enumerate}
\usepackage{epsfig}
\usepackage{amssymb}
\usepackage{amsmath}
\usepackage{indentfirst}
\usepackage{chngcntr}
\usepackage{color}
\journal{Journal of \LaTeX\ Templates}

\usepackage{graphics}
\biboptions{sort&compress}

\usepackage{graphicx}

\begin{document}
\sloppy

\begin{frontmatter}

\title{Experimental determination of the Dalitz plot for positronium decay using the J-PET detection system
}





\author[IPJ,TC]{Magdalena~Skurzok} 
\author[IPJ,Kitz]{Steven~D. Bass}  
\author[IPJ,TC]{Kamila Kasperska}
\author[IPJ,TC]{Ermias Beyene}
\author[IPJ,TC]{Neha~Chug}  
\author[INFN]{Catalina~Curceanu}  
\author[IPJ,TC]{Eryk~Czerwiński}
\author[IPJ,TC]{Manish~Das} 
\author[MSC]{Marek Gorgol} 
\author[IPJ,TC]{Sharareh Jalali} 
\author[MSC]{Bożena Jasińska}
\author[IPJ,TC]{Krzysztof~Kacprzak} 
\author[IPJ,TC]{Tevfik Kaplanoglu}
\author[IPJ,TC]{Łukasz~Kapłon}  
\author[IPJ,TC]{Aleksander Khreptak} 
\author[IPJ]{Tomasz Kozik}  
\author[IPJ,TC]{Deepak Kumar} 
\author[IPJ,TC]{Karol Kubat} 
\author[IPJ,TC]{Sumit Kumar Kundu} 
\author[CUT]{Edward Lisowski} 
\author[CUT]{Filip Lisowski} 
\author[IPJ,TC,AGH]{Bartłomiej~Łach} 
\author[IPJ,TC]{Justyna Mędrala-Sowa} 
\author[IPJ,TC]{Wiktor Mryka} 
\author[IPJ,TC]{Simbarashe Moyo} 
\author[IPJ,TC]{Szymon Niedźwiecki} 
\author[IPJ,TC]{Anand Pandey} 
\author[IPJ,TC]{Piyush Pandey} 
\author[IPJ,CIT,INFN,CAT]{Alessio Porcelli} 
\author[AGH]{Bartłomiej Rachwał} 
\author[IPJ,TC]{Martin Rädler} 
\author[IPJ,TC]{Sushil Sharma} 
\author[IPJ,TC]{Ewa Stępień} 
\author[AGH]{Tomasz Szumlak} 
\author[IPJ,TC]{Pooja Tanty} 
\author[IPJ,TC]{Keyvan Tayefi Ardebili} 
\author[IPJ,TC]{Satyam Tiwari} 
\author[IPJ,TC]{Kavya Valsan Eliyan} 
\author[IPJ,TC]{Anoop K. Venadan} 
\author[IPJ,TC]{Paweł~Moskal}

\address[IPJ]{Faculty of Physics, Astronomy and Applied Computer Science, Jagiellonian University, prof.\ Stanis{\l}awa {\L}ojasiewicza~11, 30-348 Krak\'{o}w, Poland}

\address[TC]{Center for Theranostics, Jagiellonian University, Kopernika 40, Kraków, Poland}

\address[Kitz]{Kitzbühel Centre for Physics, Kitzbühel, Austria} 


\address[INFN]{INFN, Laboratori Nazionali di Frascati, Via E. Fermi, 40, 00044 Frascati (Roma), Italy} 
 
\address[MSC]{Institute of Physics, Maria Curie-Sklodowska University, Lublin, Poland} 

\address[AGH]{AGH University of Krakow, al. Adama Mickiewicza 30, 30-059 Kraków, Poland}

\address[CUT]{Faculty of Mechanical Engineering, Cracow University of Technology, 37 Jana Pawła II Av., 31-864 Kraków, Poland}

\address[CIT]{Centro de Investigación, Tecnología, Educación y Vinculación Astronómica, Universidad de Antofagasta,
Avenida Angamos 601, 1240000 Antofagasta, Chile}

\address[CAT]{Center of Astronomical Research, Technology, Education, and Outreach, University of Antofagasta, Avda. U. de Antofagasta 02800, 1240000 Antofagasta, Chile}

\begin{abstract}

We present the first measurements of the Dalitz plot 
for ortho-positronium annihilation to three photons.
Our measurements, 
accurate to about 3\% statistical and 2-3\% systematic uncertainty in angular representation over almost the entire available phase space, 
were performed using the Jagiellonian Positron Emission Tomograph (J-PET) based on organic scintillator strips.
Until now,  
the Dalitz plot for the three-body positronium decay has been poorly explored. 
The new measurements presented here are consistent 
with both the leading-order and next-to-leading order QED predictions for the Dalitz plot.

\end{abstract}

%


\end{frontmatter}

\section{Introduction}

Studies of positronium are focused on measurements of its decays and spectroscopy.
For positronium in vacuum these observables are described by QED bound-state theory. The predictions are being tested up to and beyond next-to-leading order corrections in the fine structure constant $\alpha$
\cite{Adkins_PhysRep2022,Cassidy_EPJ2018,SBSMPMES,StevenBass}.
%
Recent theoretical developments in the context of spin-zero parapositronium are reported in Refs.~\cite{Czarnecki:2026ect,Piotrowska:2024nfj}.
%
In addition, 
positronium is being used to perform new tests of QED discrete 
symmetries 
\cite{Moskal_Nature2021,Moskal_Nature2024,J-PET:2025xie,Cassidy:2025}
which go beyond the tests in single-electron systems.
Beyond measurements in vacuum, 
positronium decays in medium play an important role in bio-medical and material sciences applications \cite{SBSMPMES,Moskal_NatRevPhys2019}. 
A highlight is the new technique of positronium imaging 
\cite{Moskal_SciAdv2021, Moskal_SciAdv2024,Moskal_IEEE2019,Das_IEEE2026, Das_SciRep2026, Kubat_JMP2026, Kasperska_BAMS2025}
that is being pioneered in medical diagnostics including the use of quantum entanglement of the final state photons resulting from the decay \cite{Moskal_SciAdv2025,Moskal_BAMS2026}. 
For the three-photon decay of the longest-lived spin-one state ortho-positronium, an essential ingredient is the Dalitz plot which encodes information about the momentum dependence of the final state photons and connects to key observables like the energy spectrum and decay rate. 
The Dalitz plot so far has not been measured in previous experiments due to 
limited kinematical coverage. 
Here we report the first measurements using the J-PET tomograph detector system in Cracow. 
Our result shows the potential of the experimental setup. 
We presently determine the Dalitz plot up to about 3\% statistical uncertainty and 2-3\% systematic uncertainty over almost the full phase space. Planned measurements with an upgraded detector system will allow about an order of magnitude improvement in precision.

This paper is structured as follows. Section 2 provides an overview of the Dalitz plot in the context of o-Ps decays. Section 3 introduces the Jagiellonian Positron Emission Tomograph (J-PET) detection system. The subsequent section outlines the experimental methodology, followed by Section 5, which presents the results. The final section presents the conclusions and perspectives for future measurements.


\section{Dalitz Plot for $o$-$Ps \rightarrow 3\gamma$ decay}

Positronium can be formed in two ground states: para-positronium, $p$-$Ps$ (the singlet state with spin zero and lifetime in vacuum of 0.125~ns) and ortho-positronium, $o$-$Ps$ (the triplet state with spin one and lifetime of 142~ns).
It is an eigenstate of parity (P), charge (C) and thus CP operator which makes it ideal for studies of the discrete symmetries in quantum electrodynamics (QED)~\cite{Sozzi,Moskal_APPB}. Charge conjugation symmetry means that $p$-$Ps$ and $o$-$Ps$ decay into even and odd numbers of photons, respectively.

In contrast to the two-photon back-to-back para-positronium decay, 
the longer-lived
ortho-positronium decays into three photons with a 
more complex phase space. In the three-photon decay process, energy-momentum conservation requires the total photon energy to be equal to the ortho-positronium mass.
That is
$E_{1}+E_{2}+E_{3}=2m_{e}$, 
where $E_{1},E_{2},E_{3}$ are the energies of the three emitted photons and $m_e$ is the electron mass 
\footnote{The sum of the three photon energies is slightly below $2 m_e$ since positronium comes with a small binding energy 
$E_{B} \approx$ -6.8eV~\cite{Cassidy_EPJ2018}}. 
The photon momenta lie in a plane $\vec{k}_{1}+\vec{k}_{2}+\vec{k}_{3}=0$ so that 
the four-momentum of any photon is completely determined by the momenta of the other two.
The photon energies take values from 0 to 511 keV. 
The differential cross-section of the $o$-$Ps \rightarrow 3\gamma$ decay in the centre-of-mass frame is determined from 
the low energy limit of the cross section for the $e^+ e^- \to 3 \gamma$ process


\begin{equation}
d\sigma_{3\gamma}=\frac{(2\pi)^{4}|M|^{2}}{4m_{e}^{2}v}\delta(\vec{k}_{1}+\vec{k}_{2}+\vec{k}_{3})\delta(E_{1}+E_{2}+E_{3}-2m_{e})
\frac{1}{(2 \pi)^9}
\frac{d^{3}k_{1}}{2E_1}
\frac{d^{3}k_{2}}{2E_2}
\frac{d^{3}k_{3}}{2E_{3}}.
\end{equation}
%
%
Here 
$v$ is the electron positron relative velocity
and the $\delta$ functions express energy and momentum conservation in the decay. 
%
The factor $|M|^2$ denotes the squared absolute value of the Lorentz invariant amplitude for the \mbox{$o$-$Ps$ $\rightarrow$ 3$\gamma$} transition summed over the final state spins of the outgoing photons and averaged over the initial state polarization of the electron and positron. 
For the lowest-order (LO) decay process (involving 6 Feynman diagrams)
$|M|^{2}$ is
\cite{OrePowel_1949,Berestetskii}:
%
\begin{equation}
|M|^{2}=16(4\pi)^3\alpha^{3}\left[\left(\frac{m_{e}-E_{1}}{E_{2}E_{3}}\right)^{2}+\left(\frac{m_{e}-E_{2}}{E_{1}E_{3}}\right)^{2}+\left(\frac{m_{e}-E_{3}}{E_{1}E_{2}}\right)^{2}\right].
\end{equation}
%
When the calculations are extended to include 
next-to-leading-order corrections at $O(\alpha)$, 
$|M|^2$ takes a much more complicated form;
for details see Ref.~\cite{Adkins_PhysRev2005}. 
Due to the energy-momentum conservation, the massless nature of the emitted photons and rotational symmetry of the decay, 
the kinematics of the three photons originating from $o$-$Ps$ decay is determined entirely by specifying only two independent photon parameters, such as the energies or relative angles between photons. 
This leads to simplification of the differential annihilation cross-section formula, which in energy representation can be expressed as:  
\begin{equation}
d\sigma_{3\gamma}=\frac{1}{6m_{e}^{2}v}\frac{1}{16(2\pi)^3}|M|^{2} dE_{1} dE_{2}.
\end{equation}\label{eq_3}
%
%
%
%
Here, 
the factor of six in the denominator refers to 6=3! possible permutations of three identical photons. 
The $o$-$Ps$ $\to 3 \gamma$ differential decay rate  
is determined
by 
averaging the cross-section over initial spin states ($\frac{4}{3} d\sigma_{3\gamma}$, three of the four possible positronium spin states undergo 3$\gamma$ annihilation), multiplying by the flux density
$v|\psi(0)|^{2}=\frac{m^{3}\alpha^{3}}{8\pi}$ 
where $|\psi(0)|^{2}$ is the wave function of the positronium evaluated at the origin,
and finally taking the limit that $v \to 0$ \cite{Berestetskii}.

\begin{figure}[t!]
\centering
\includegraphics[width=15.0cm,height=7.0cm]{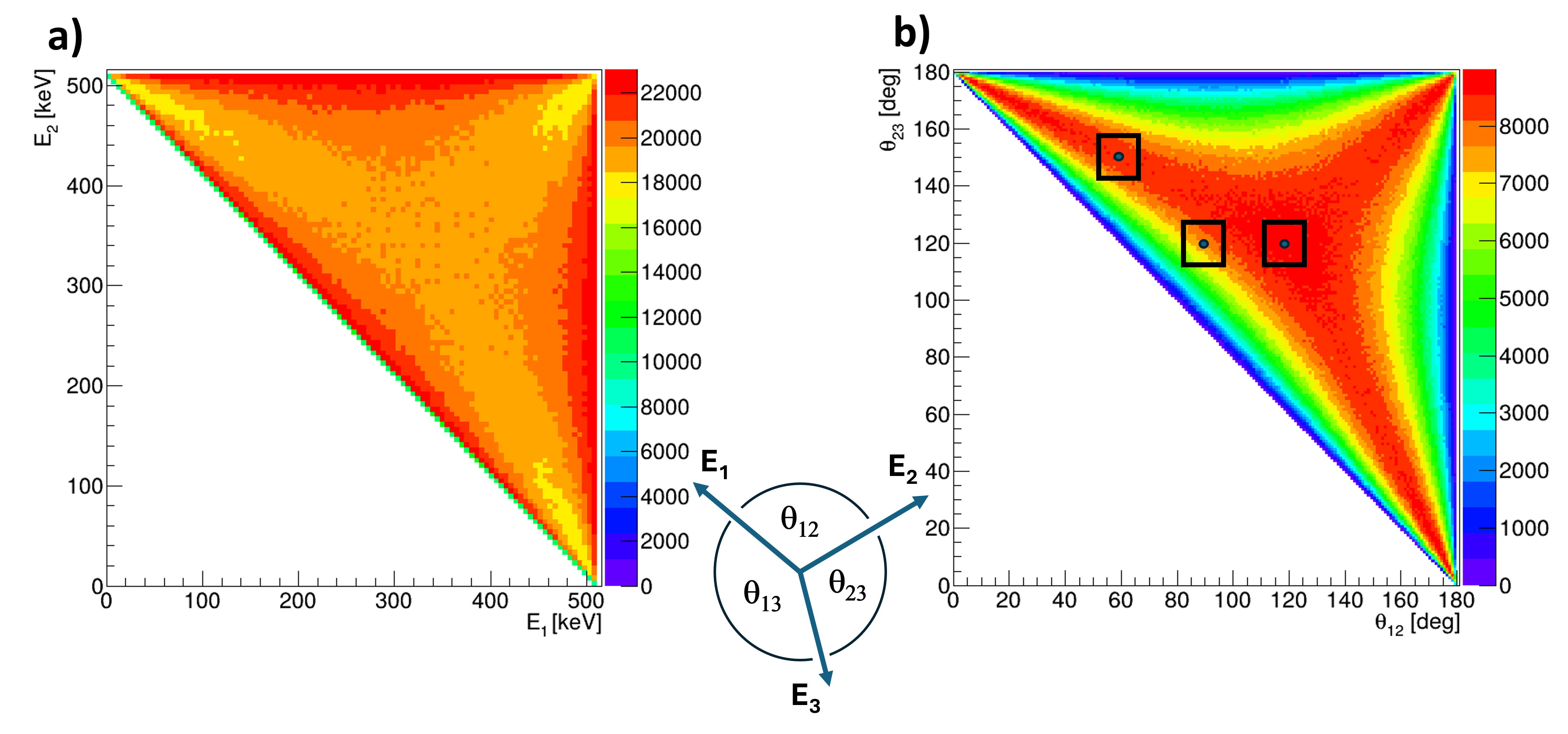} 
\caption{Energy (a) and corresponding angular (b) distribution for $o$-$Ps \rightarrow 3\gamma$ decay. The distribution was generated for $10^{8}$ events based on the QED predictions for LO decay process~\cite{OrePowel_1949,Berestetskii}. The black rectangles denote three angular configurations measured in  Ref.~\cite{MillsBerko}. 
\label{fig_dalitz_gen}}
\end{figure}

The complete kinematics of the $o$-$Ps \rightarrow 3\gamma$ process can be vizualised using a Dalitz plot, where the axes correspond to two independent observables of the emitted photons. This excellent tool introduced by Dalitz~\cite{Dalitz} has been widely and successfully applied in nuclear and particle physics for many years to study three-body decays, 
see  e.g.~\cite{Amsler:1997up,WASA-at-COSY:2008rsh,DalitzPlot2,DalitzPlot3,DalitzPlot4}.
Each point in the Dalitz plot corresponds to a specific kinematic configuration of the three photons and is weighted by the squared QED matrix element $|M|^{2}$ so that the event density directly reflects the differential decay rate. 
The total decay rate is recovered by integration over the full kinematically allowed region, taking into account the Bose symmetry of the photons and the associated phase-space constraints.
As an example, 
the Dalitz plot of $o$-$Ps$ decay into three photons in energy and angular representation generated based on LO QED calculations is presented in Fig.~\ref{fig_dalitz_gen}.

If we can measure the complete Dalitz plot for the three-photon $o$-$Ps$ decay, 
this would provide a multidimensional view of the $o$-$Ps \rightarrow 3\gamma$ decay dynamics, enabling sensitive tests of differential QED amplitudes (differential decay rates) beyond traditional observables such as total decay rates.
The Dalitz plot for 
ortho-positronium decays has never been measured in the entire available phase space. 
So far, only three possible angular configurations of annihilation photons (marked in Fig.~\ref{fig_dalitz_gen}b)) have been measured with limited statistics~\cite{MillsBerko}. For each of these configurations, the measurement provided only
the total number of counts within an angular range of $\pm$8.5$^{\circ}$ (marked as black rectangles in Fig.~\ref{fig_dalitz_gen}b))  resulting from the geometry of the detector used, that is, using crystal diameters of 3.8 cm 
placed at a 
distance of 12.7 cm 
from the source.
%
The first measurements of the Dalitz plot for $o$-$Ps \rightarrow 3\gamma$ annihilation process in almost entire available kinematic region are presented 
below in Section 4.
This experiment was 
performed by the J-PET group in Kracow with a multiphoton detector system based on organic scintillators. 

\begin{figure}[t!]
\centering
\includegraphics[width=8.0cm,height=7.0cm]{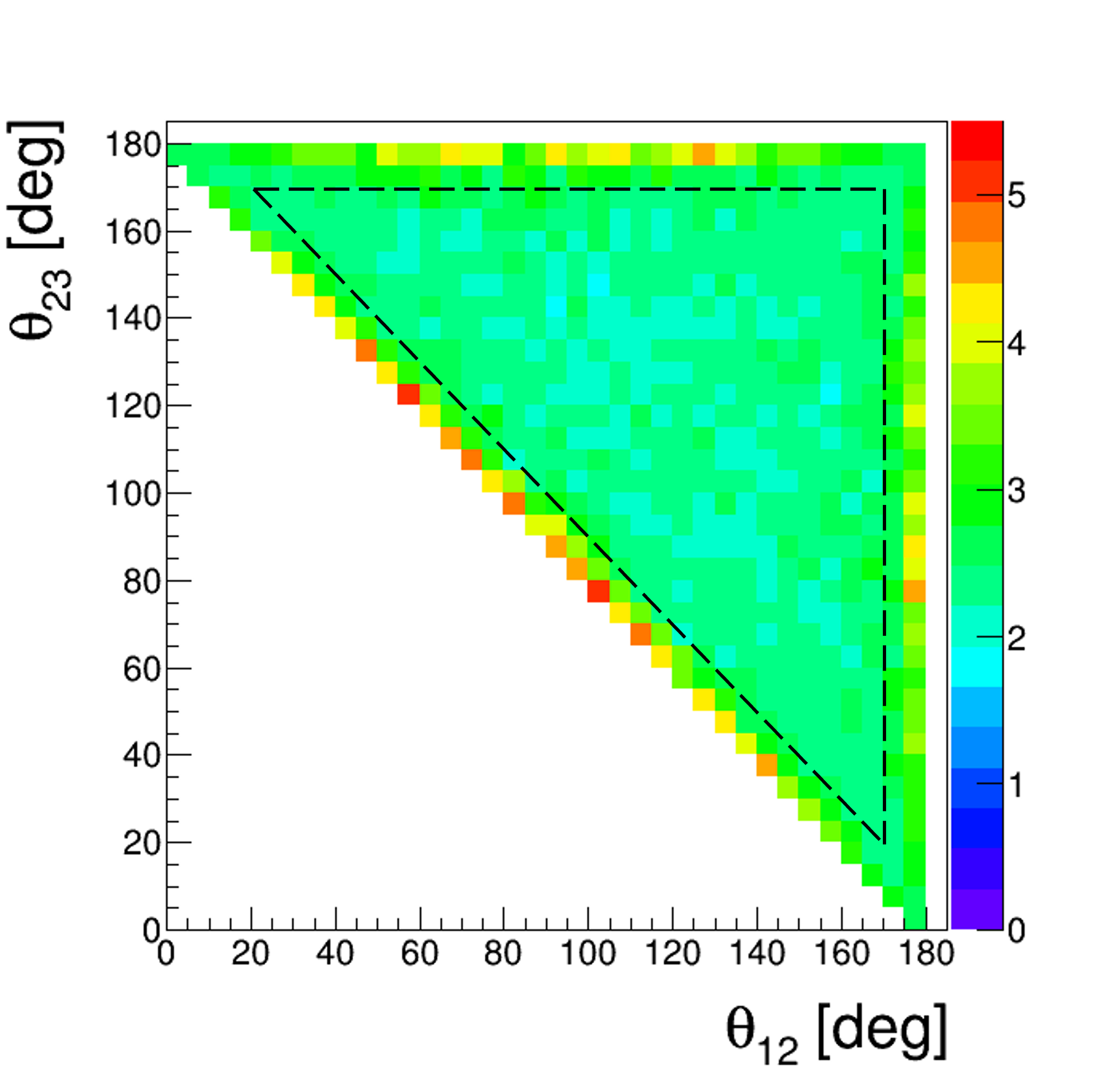} 
\caption{The relative percentage difference between the QED leading-order (LO) prediction and LO + O($\alpha$) result, normalized to the LO prediction, in the angular distribution for the $o$-$Ps \rightarrow 3\gamma$ decay ($\frac{N_{\rm LO}-N_{\rm LO+ O(\alpha)}}{N_{\rm LO}}\cdot 100\%$). The distribution was obtained based on simulations including $|M|^{2}$ for the LO term and O($\alpha$) correction, as described in~\cite{Adkins_PhysRev2005}. The black lines indicate the experimental limits of J-PET (see Sec.~\ref{Results}). In the region of interest (marked with the dashed line)
the difference between LO and LO+O($\alpha$) correction varies between 1.5 - 2.5\%.
\label{fig_dalitz_diffr}}
\end{figure}

Ultimately, as a long-term goal, one would like to measure the Dalitz plot to sufficient precision that we become sensitive to higher-order QED corrections at $O(\alpha)$ and beyond.
Experiments have verified 
the $O(\alpha)$ correction to the total decay rate 
as about 2-3\% correction to the leading term \cite{Vallery:2003iz,Jinnouchi:2003hr,Kataoka:2008hj}. 
Theoretical predictions are available to $O(\alpha^3)$, with testing the non-leading terms a challenge for future experiments.
Also, $O(\alpha)$ corrections to the energy spectrum in the decay
were found to be needed to understand the data in Ref.~\cite{Adachi2015}.
%
Positronium spectroscopy is also a topic of vigorous investigation
\cite{Adkins_PhysRep2022}.
%
%
Positronium energy levels have been calculated to order 
$m_e \alpha^6$ 
(plus some terms to higher order)
with the leading term starting at ${O} (m_e \alpha^4)$.
Recent transition frequency experiments are reported in Refs.~\cite{Sheldon:2023vgh,Gurung_PhysRevLett2020,Gurung_PhysRevA2021,Borges:2025qfk}.
While these experiments are mostly in good agreement with QED theory, 
there is also a more than 4 standard deviations 
(or one part in $10^4$) 
anomaly in the $2 ^3S_1 \to 2 ^3P_0$ transition \cite{Gurung_PhysRevLett2020,Gurung_PhysRevA2021}.
%
%
Eventually, one would like to connect especially the decay rate and energy spectrum measurements with the Dalitz plots. 
%
%
%
So far,  
higher-order effects, in particular O($\alpha$) corrections in the angular distribution have not been investigated experimentally. Fig.~\ref{fig_dalitz_diffr} shows the relative difference between the leading order QED  calculation and 
the calculation 
\cite{Adkins_PhysRev2005,Adachi2015} where 
$O(\alpha)$ corrections are included.
%
If one could measure to this sensitivity, 
then this detailed characterization would allow to test higher-order effects, in particular O($\alpha$), provide a reference dataset for future theoretical studies, and put much tighter constraints on three-body annihilation amplitudes.

\section{The J-PET experimental setup}

The J-PET apparatus~\cite{Moskal_NIMP2014,Niedzwiecki_APPB2017}, located at the Jagiellonian University in Kracow, Poland, is a unique and innovative detection system developed based on plastic scintillation material for simultaneous multi-photon measurements. This feature makes it an excellent tool for studying positronium atom decays, in both, fundamental research~\cite{Moskal_Nature2024,Moskal_Nature2021} and medical imaging applications~\cite{Moskal_SciAdv2024,Moskal_SciAdv2021,Das_IEEE2026,Das_SciRep2026,Kubat_JMP2026,Das_BioAlg2025,Satyam_BAMS2025,Radler_BAMS2025}.

\begin{figure}[h]
\centering
\includegraphics[width=15.5cm,height=9.5cm]{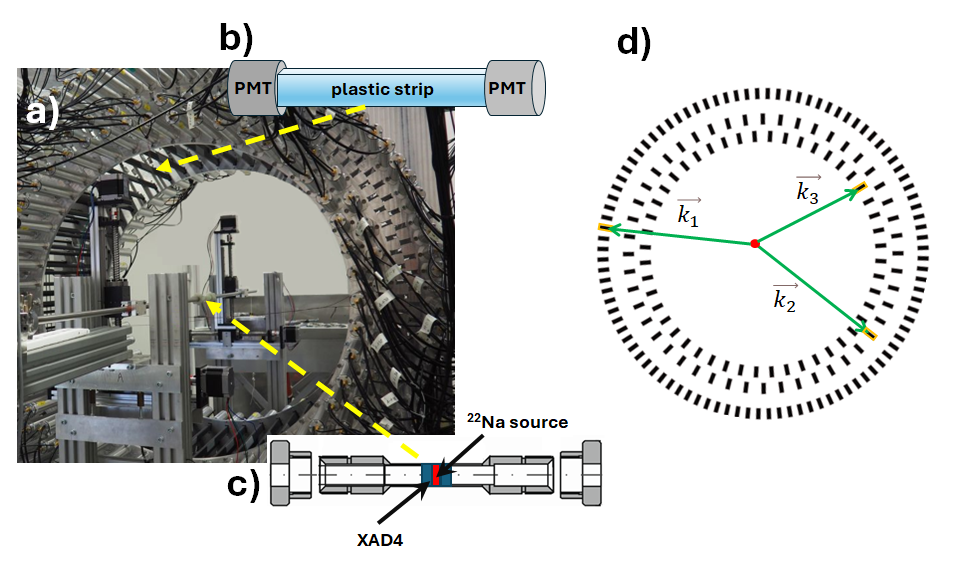}
\caption{(a) Photograph of the J-PET detector built from 192 strips of EJ-230 scintillators, each read out by R9800 Hamamatsu PMTs~\cite{Niedzwiecki_APPB2017} (b); with a small annihilation chamber containing a sodium $^{22}\hspace{-0.03cm}\mbox{Na}$ radioactive $\beta^{+}$ embedded in porous XAD-4 material, 
fixed in the center~\cite{Gorgol_APPB2020} (c). (d) Schematic of the J-PET detector showing photons (green arrows) associated with the $o$-$Ps \rightarrow 3\gamma$ annihilation occurring inside the chamber (red dot). \label{fig_2}}
\end{figure}

The analysis presented here relies on measurements conducted with the J-PET detector using a small annihilation chamber as a positronium target (see Fig.~\ref{fig_2}). 
The J-PET detector consists of 192 plastic scintillator strips, each 500 mm long, 7 mm thick, and 19 mm high, arranged in three concentric cylindrical layers (Fig.~\ref{fig_2} (a)). Each of the two inner layers consists of 48 strips and is rotated by 3.75$^\circ$ with respect to the other. These layers are positioned at radii of 42.5~cm and 46.75~cm, respectively. The outer layer, located at a radius of 57.5 cm, is composed of 96 strips. Each scintillator strip is wrapped in two types of foils: an outer foil for optical isolation and an inner reflective foil to enhance light collection in R9800 Hamamatsu photomultipliers (PMTs) optically connected to both its ends (Fig.~\ref{fig_2} (b)). The light is generated as a result of photons undergoing Compton scattering in the scintillator material and converted into electrical signals by PMTs. The J-PET data acquisition system (DAQ), based on Field Programmable Gate Arrays (FPGAs) operates in a trigger-less mode~\cite{Korcyl_IEEE2018,Palka_JINST2017}. The analog signals from the PMTs are probed at four amplitude thresholds: 30, 80, 190, and 300 mV, allowing for up to eight time-point measurements (four at the leading edge and four at the trailing edge), which are subsequently digitized by a Time-to-Digital Converter (TDC) and distributed to the Trigger Readout Boards (TRBs). To ensure consistent and precise timing measurements across the entire detector system, proper time synchronization is performed~\cite{Kacprzak_BAMS2026}.
The longitudinal position of the photon
interaction in the strip is reconstructed based on the difference in the arrival times of light signals at the two ends, while the interaction time is determined as the average of the two times \cite{Moskal_NIMP2014}.
Fast timing response of the plastic scintillators allows one to achieve an interaction time resolution of approximately 250~ps, while an angular resolution reaches about~1$^\circ$~\cite{Moskal_Nature2021}.


For efficient production of positronium atoms, a properly designed annihilation chamber was used and positioned at the centre of the J-PET detector (Fig.~\ref{fig_2}(a,c))\cite{Jasinska_APPB2016,Gorgol_APPB2020}. The chamber consists of an internal pipe containing a $^{22}\hspace{-0.03cm}\mbox{Na}$ positron source with activity of 0.7~MBq, surrounded by a porous XAD-4 polymer, and placed inside a larger, external cylindrical chamber. The XAD-4 polymer is characterized by a high fraction of positron-electron 3$\gamma$ annihilations (29\%) and a high $o$-$Ps$ formation probability~\cite{Jasinska_APPB2016,Gorgol_APPB2020}. The external chamber, 1~mm thick and made of lightweight PA6 polyamide (1.14~g/cm³), attenuates annihilation photons by only 1\%. The chamber is connected to vacuum pump system (two rotary pumps) by steel pipes, which also set the whole chamber in the axis of J-PET detector. The cylinder is closed from both ends with the plastic partitions that allows to reach the vacuum at the order of~1.5$\cdot$10$^{-5}$ Pa. 
%
%
The measurements reported here are based on 
data from 218 effective days of measurements with the sodium source resulting in the identification of 33 million $o$-$Ps \rightarrow 3\gamma$ signal events. The volume of processed data amounts to approximately 950~TB.

\section{Selection of $o$-$Ps \rightarrow 3\gamma$ annihilation event candidates}

In the experiment, interactions of positrons produced in the decay of a sodium source ($^{22}\hspace{-0.03cm}\mbox{Na}$ $\rightarrow$ $^{22}\hspace{-0.03cm}\mbox{Ne}^{*}$ $e^+ \nu_{e}$ $\rightarrow$ $^{22}\hspace{-0.03cm}\mbox{Ne}$ $\gamma_{d} e^+ \nu_{e}$) with electrons in the porous material lead to annihilation processes to two- or three-photons. 
The annihilation may occur directly or via the formation of a positronium atom and is preceded by the emission of a prompt photon ($\gamma_{d}$) from the excited daughter nucleus ($^{22}\hspace{-0.03cm}\mbox{Ne}^{*}$), having energy of 1275~keV. 

To extract three-photon $o$-$Ps$ decays from the experimental data, appropriate selection criteria were applied. These were primarily based on Monte Carlo simulations. The dedicated simulations were performed using GEANT4 simulation package~\cite{Alison_PRS2016,Alison_IEEE2006,Agostinelli_NIM2003} with implemented geometries of J-PET tomograph and annihilation chamber, and material composition. The fraction of direct positron-electron annihilations, $p$-$Ps$ and $o$-$Ps$ formations known for the XAD-4 porous material were used. The simulated data, once convoluted with the detector response, were then analysed with dedicated framework~\cite{Moskal_SciAdv2024} using the same procedure as for the experimental data.

\begin{figure}[b!]
\centering
\includegraphics[width=15.0cm,height=6.0cm]{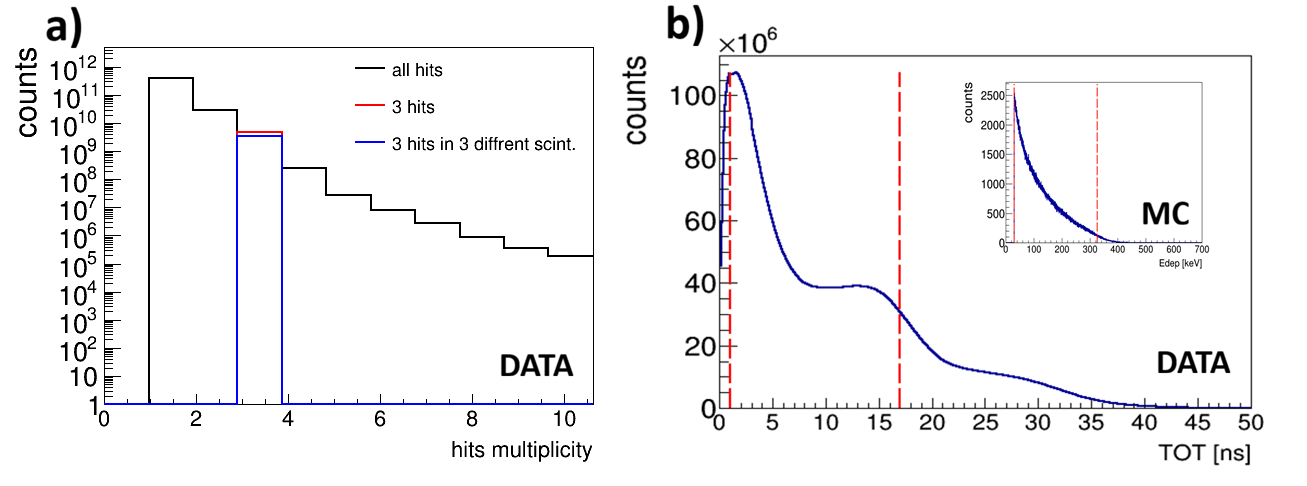}  \hspace{-0.4cm}
\caption{(a) Experimental hits multiplicity before and after applying the requirement of the strip’s active region and the condition of three interactions in three different scintillators.
(b) Experimental TOT distribution related to energy deposition in the scintillators. The red dashed line on the right indicates the Compton edge corresponding to 511 keV photons from 2$\gamma$ annihilation, while the line on the left corresponds to the energy threshold set in the electronics. The corresponding energy deposition distribution obtained from MC simulations of annihilation into 3$\gamma$ for the signal is shown in the inset plot.
\label{fig_cuts1}}
\end{figure}

In the first step of selection, the accepted events corresponded 
to the registration of three photon hits in the active regions of three different scintillator strips (hits located within a $\pm$23 cm around the strip's center), within a 2.5~ns coincidence time window 
(see Fig.~\ref{fig_cuts1}(a)). 
To distinguish between photons from different origins, i.e., annihilation and de-excitation photons, the Time-over-Threshold (TOT) technique was applied~\cite{Sharma_EJNMMI2020}. TOT, which is related to the energy deposited by photons in plastic scintillators, was determined for each hit as the sum of signal widths at four predefined voltage thresholds~\cite{Sharma_EJNMMI2020,Kacprzak_BAMS2026}. The TOT spectrum for three-photon events is shown in Fig.~\ref{fig_cuts1}(b). For further analysis, a TOT range from 1 to 17 ns was selected (area between red dashed vertical lines), where a contribution from annihilation photons is expected. 


\begin{figure}[t!]
\centering
\includegraphics[width=16cm,height=6.0cm]{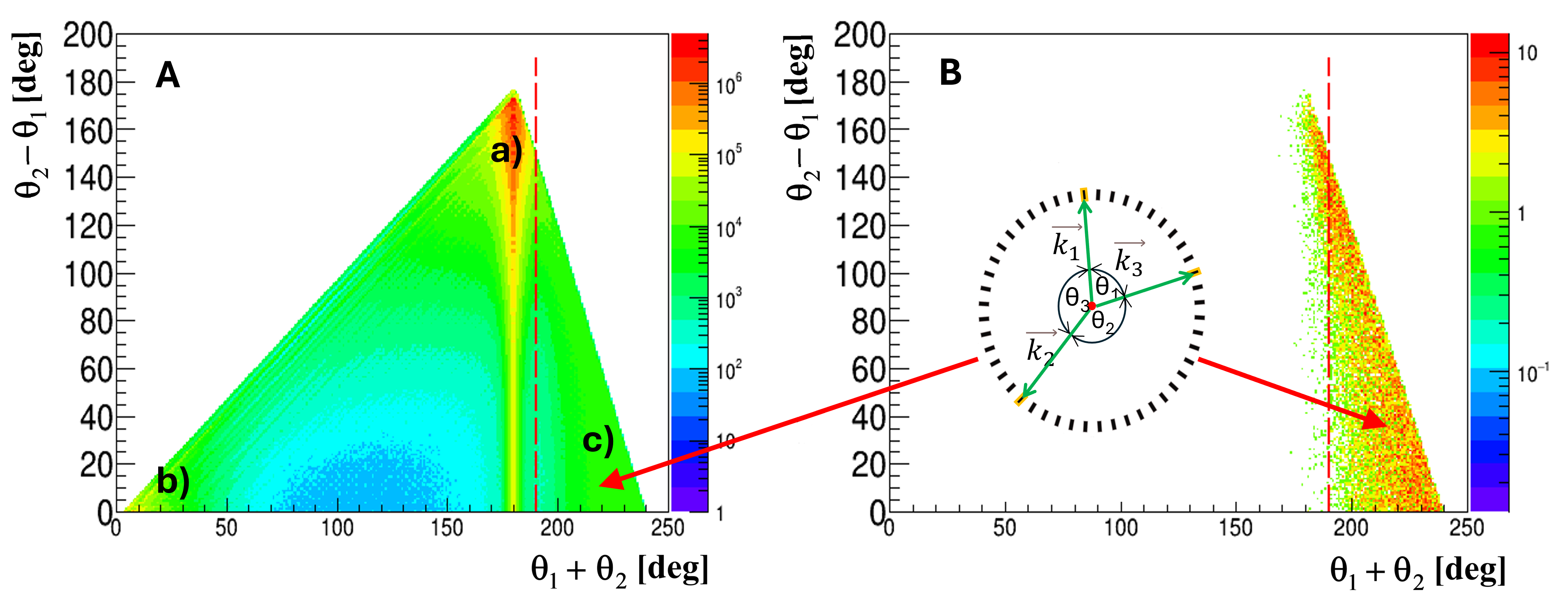} 
\includegraphics[width=14.0cm,height=5.0cm]{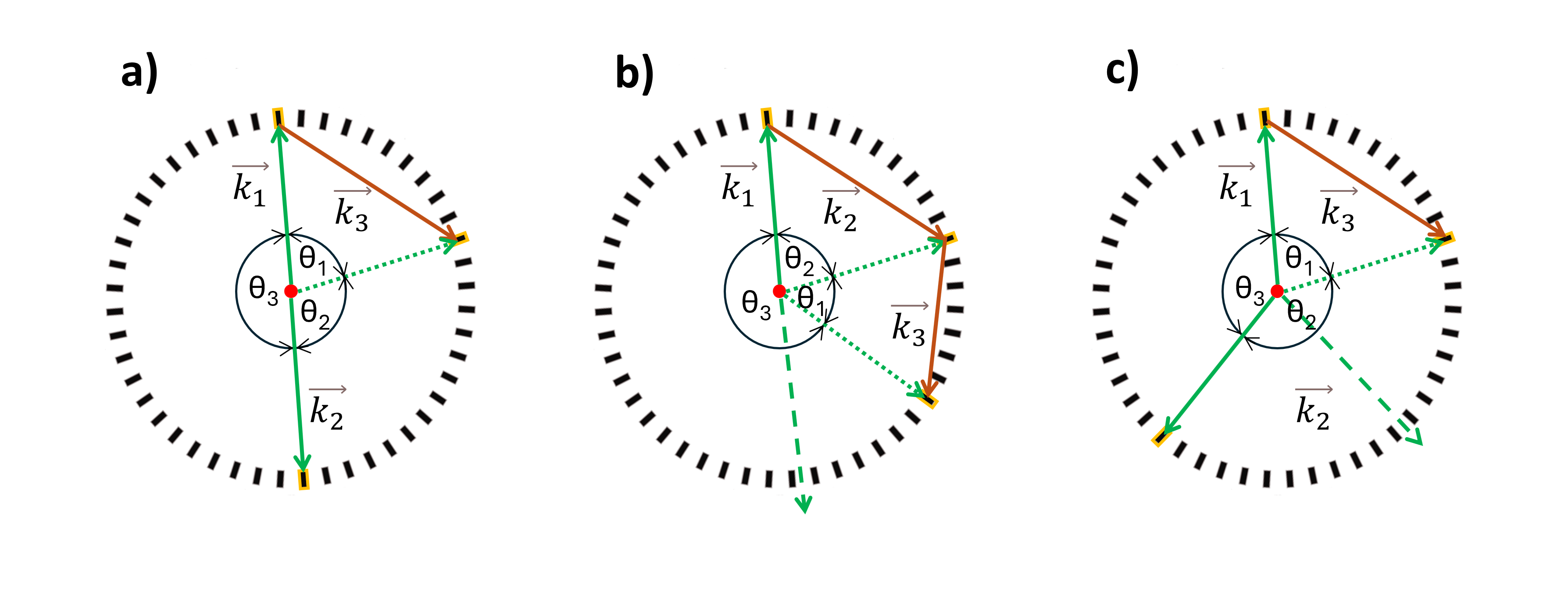}  
\caption{Spectrum of $\theta_{2}-\theta_{1}$ vs. $\theta_{2}+\theta_{1}$ for 3$\gamma$ events being candidates for $o$-$Ps \rightarrow 3\gamma$ decays for experimental data (Fig. A) and obtained from simulations of the signal (Fig. B). Lower panel represents main background contributions corresponding to regions a), b) and c) shown in the upper panel. The background contributions are: a) 2$\gamma$ annihilation with secondary Compton scattering, b) two back-to-back photons, where one photon is not registered and the other undergoes double scattering, c) misreconstructed $o$-$Ps \rightarrow 3\gamma$ decay. Registered annihilation photons are denoted by solid green arrows, not registered annihilation photons by the dashed green arrows, while wrongly assigned photons by the dotted green arrow.} 
\label{fig_cuts3}
\end{figure}

\begin{figure}[t!]
\centering
\includegraphics[width=7.5cm,height=6.0cm]{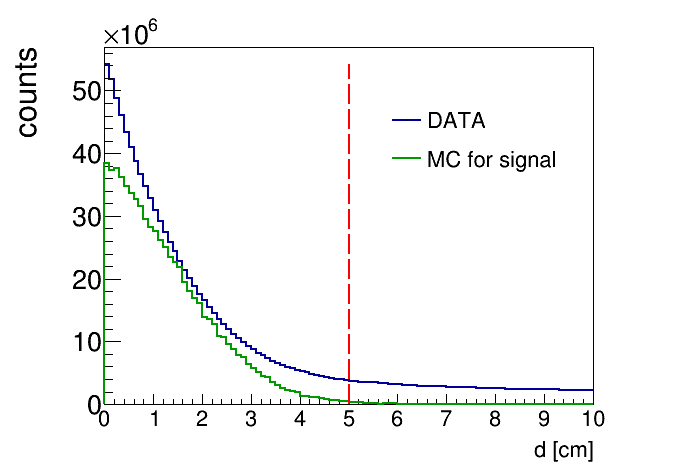}  \hspace{-0.4cm}
\includegraphics[width=7.5cm,height=6.0cm]{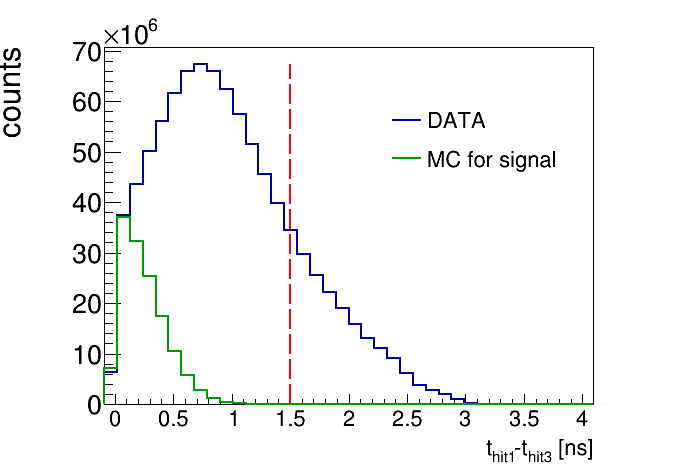}  
\caption{(left) Distance between annihilation origin and the decay plane. (right) Emission time difference between the first and the third hit in event. Applied cuts $d$<5cm and $t_{\rm hit1}-t_{\rm hit3}$<1.5ns are marked with red dashed vertical lines. 
\label{fig_cuts2}}
\end{figure}

To substantially reduce background originating from secondary scattering of $\gamma$ quanta (see Fig.~\ref{fig_cuts3}a,b), cosmic rays, and random coincidences, selection criteria were applied based on the distance $d$ between the decay plane and the annihilation point (defined by the position of the annihilation chamber), as well as the emission time difference between the first and last hit in the cluster $\delta t_{13}=t_{\rm hit1}-t_{\rm hit3}$ (for details see \cite{Moskal_SciAdv2024,Moskal_SciAdv2025}). The corresponding distributions are shown in Fig.~\ref{fig_cuts2}. For further analysis, events corresponding to $d$<5~cm and $\delta t_{13}$<1.5~ns were accepted.

The remaining background associated with single- and double-scattered events (schematically shown in Fig.~\ref{fig_cuts3}a and Fig.~\ref{fig_cuts3}b) was suppressed using angular correlations between photon momenta. Fig.~\ref{fig_cuts3}A and Fig.~\ref{fig_cuts3}B present the difference between the two smallest relative angles ($\theta_{2}-\theta_{1}$) as a function of their sum ($\theta_{2}+\theta_{1}$) for experimental data and Monte Carlo simulations for the signal, respectively. The $o$-$Ps \rightarrow 3\gamma$ events correspond to the high $\theta_{2}+\theta_{1}$ values with a peak around $240^{\circ}$ (Fig.~\ref{fig_cuts3}B). Two-$\gamma$ annihilation followed by scattering appears as a vertical band around $180^{\circ}$, while small values of $\theta_{2}+\theta_{1}$ (around $50^{\circ}$) correspond to situation where only one photon is detected and undergoes two scatterings. The signal-rich region was selected requiring $\theta_{2}+\theta_{1}>190^{\circ}$.

The last stage of the selection was aimed at subtracting events corresponding primarily to misreconstructed $o$-$Ps \rightarrow 3\gamma$ decays. This occurs when one of the annihilation photons is not detected and instead a scattered photon is registered (or less likely when only one annihilation photon is detected along with two scattered photons originating from it) as shown in Fig.~\ref{fig_cuts3}c. 

\begin{figure}[t!]
\centering
\includegraphics[width=7.5cm,height=7.0cm]{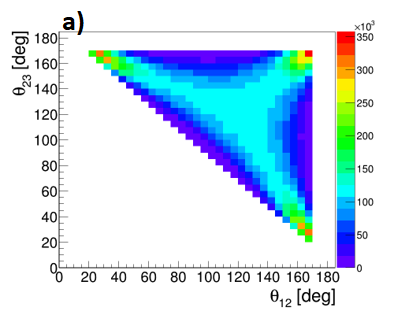} 
\includegraphics[width=7.5cm,height=7.0cm]{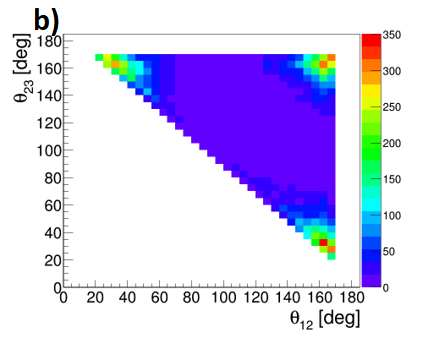}
\vspace{-0.1cm}
\includegraphics[width=15.0cm,height=7.0cm]{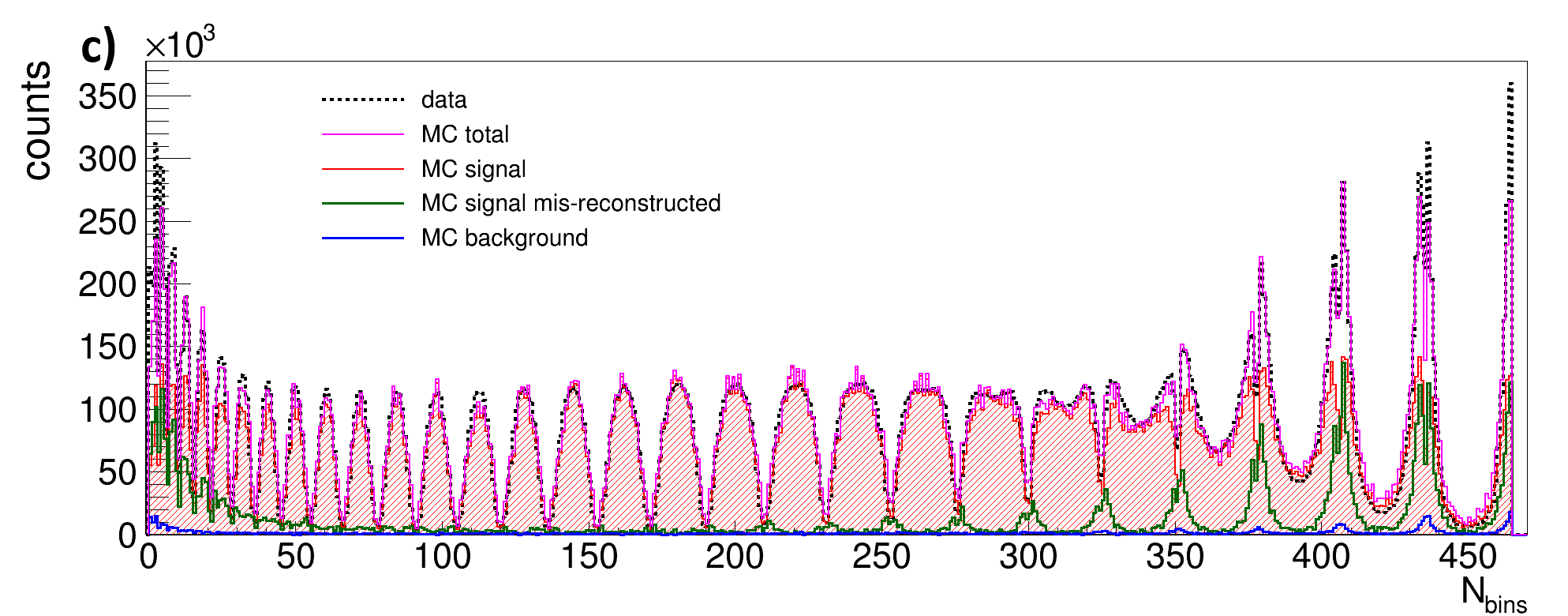}
\vspace{-0.4cm}
\caption{a) Experimental Dalitz Plot in angular representation. b) Dalitz Plot for misreconstructed signal events obtained from Monte Carlo simulations. c) The experimental Dalitz distribution (a) transferred to one-dimensional histogram, excluding empty bins. The distribution was fitted with three contributions: $o$-$Ps$ signal (red shaded area), misreconstructed signal (green solid line), the background (blue solid line), and overall fit (magenta solid line) as explained in the text.
}
\label{fig_dalitz_before_clean}
\end{figure}

To extract the misreconstructed $o$-$Ps$ events and the non-signal background from the sample, the experimental Dalitz distribution in the angular representation (Fig.~\ref{fig_dalitz_before_clean}a)) was transferred to a one-dimensional histogram, $N^{\rm exp}$, (Fig.~\ref{fig_dalitz_before_clean}c). The transformation was performed by looping over the bins of the two-dimensional distribution, first along the x-axis and then along the y-axis, and assigning each non-zero two-dimensional bin content to a consecutive bin of the one-dimensional histogram. The resulting distribution was fitted with three components: reconstructed $N^{\rm sig}$ and misreconstructed $N^{\rm mis-rec-sig}$ signal, and the background $N^{\rm bcg}$.

The fit proceeds through the minimization of a chi-square function (using the TMinuit library~\cite{Minuit_ref}): 
\begin{equation}
\chi^2 = \sum\limits_{i=1}^{N^{1D}_{\rm bins}} \frac{( N_i^{\rm exp} - A \cdot N_{i}^{\rm sig} - A \cdot N_i^{\rm mis-rec-sig} - B \cdot N_{i}^{\rm bcg})^2}{N_i^{\rm exp} + A^{2} (N_{i}^{\rm sig} + N_i^{\rm mis-rec-sig}) + B^{2} \cdot N_{i}^{\rm bcg}} \ .
\end{equation}
%
Here $i$ represents the bin index, {$N^{1D}_{\rm bins}$ is the number of bins in the one-dimensional histogram shown in Fig.~\ref{fig_dalitz_before_clean}c, and $A$ and $B$ are the fit parameters for the signal (reconstructed and misreconstructed) and the background, respectively.  



As can be seen in Fig.~\ref{fig_dalitz_before_clean}b, the contribution from partially reconstructed signal is significant in the "corners" of the Dalitz distribution which corresponds to the first and last bins in Fig.~\ref{fig_dalitz_before_clean}c. A similar pattern is observed for the remaining background component. Based on the agreement between the reconstructed Monte Carlo and the experimental data, the purity and overall efficiency are estimated to be 84\% and 0.03\%, respectively.



\begin{figure}[t!]
\centering
\includegraphics[width=7.5cm,height=7.0cm]{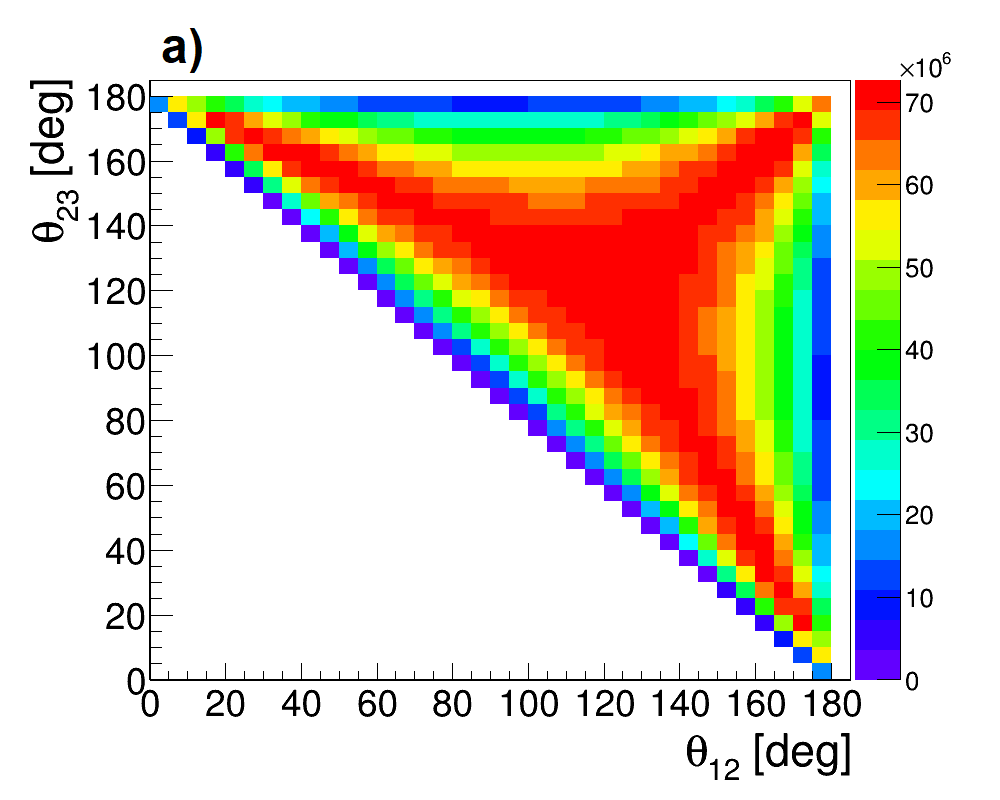}
\includegraphics[width=7.5cm,height=7.0cm]{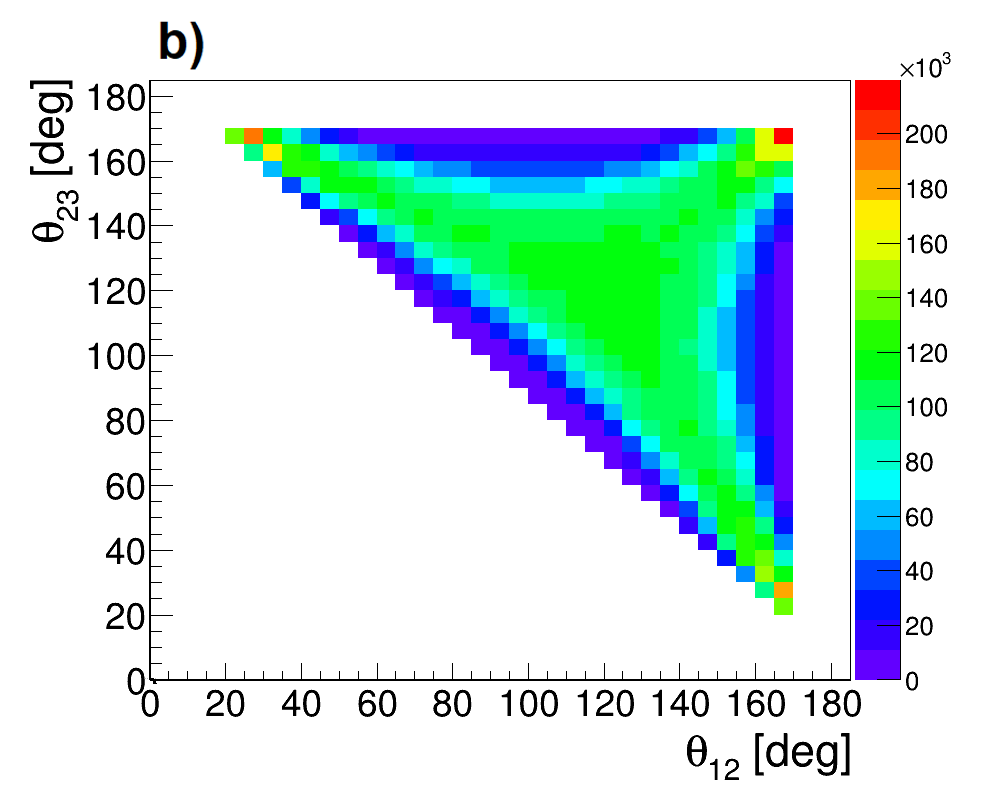}\\
\includegraphics[width=11.5cm,height=7.0cm]
{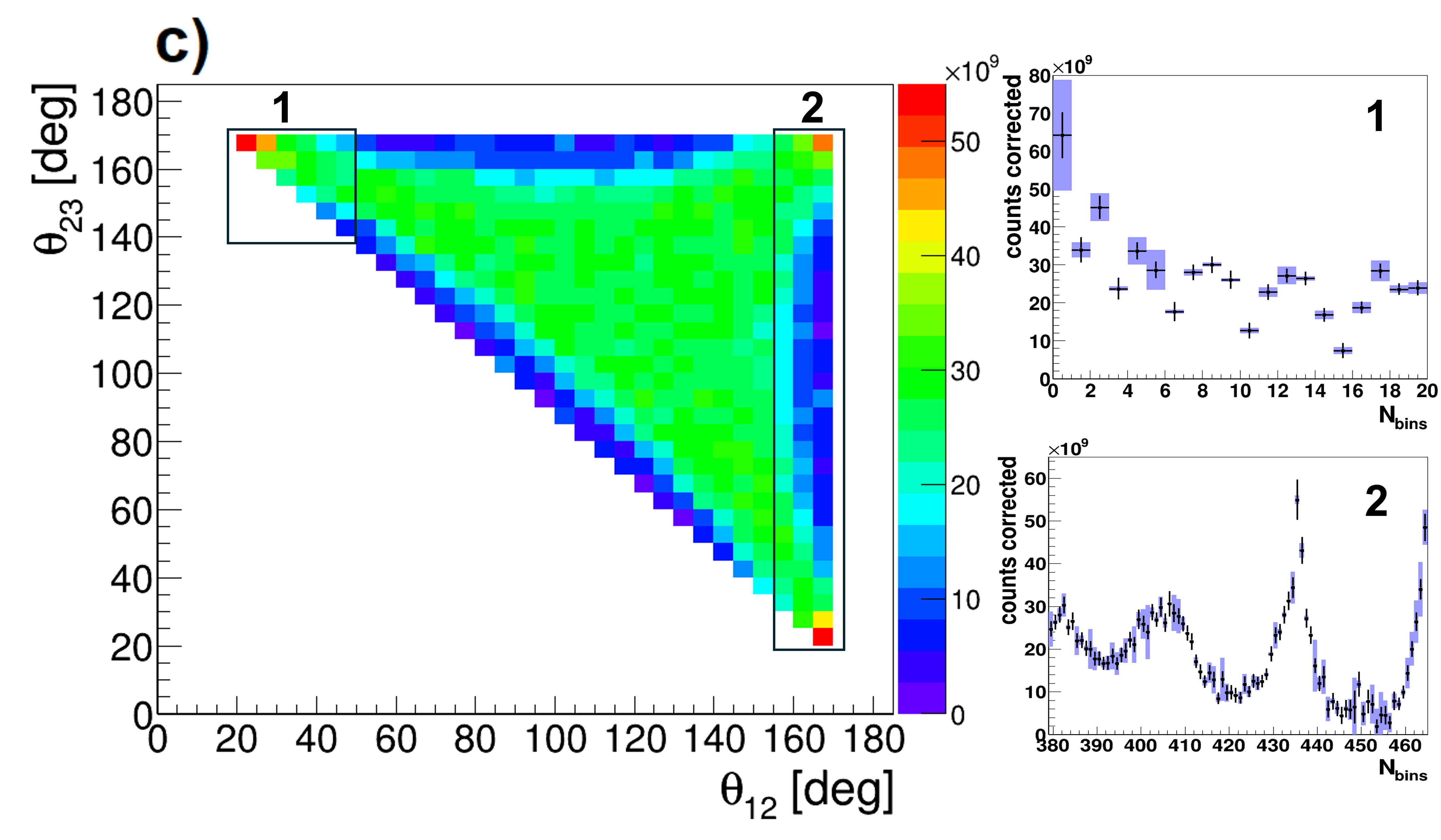}
\vspace{-0.4cm}
\caption{a) Generated Dalitz distribution based on the QED predictions for LO decay process with the same binning as that applied in experimental data analysis. b) Experimental Dalitz Plot in angular representation after background subtraction. c) The experimental Dalitz Plot (b) corrected with efficiency. The right subpanels show the systematic uncertainties for regions 1 and 2 of the Dalitz Plot.}
\label{fig_dalitz_exp_mc2}
\end{figure}

\section{Results}\label{Results}

The Dalitz distribution after subtraction of $N^{\rm mis-rec-sig}$ signal, and the background $N^{\rm bcg}$ contributions (shown in Fig.~\ref{fig_dalitz_before_clean}c) is presented in Fig.~\ref{fig_dalitz_exp_mc2}b and contains of 33.42 × 10$^{6}$ identified $o$-$Ps \rightarrow 3\gamma$ events. The distribution was corrected for overall detection and reconstruction efficiency map determined based on performed Monte Carlo simulation. The efficiency was calculated as the ratio of the number of reconstructed events to the number of generated events for each angular interval in the Dalitz spectrum. The detector response was simulated for the generated events, including the registration efficiency of the J-PET detector as a function of photon energy deposition~\cite{Sharma_EJNMMI2023}. The analysis was then performed using the same selection criteria as those applied to the experimental data.






The obtained Dalitz plot for three-photon $o$-$Ps$ decay is shown in~Fig.~\ref{fig_dalitz_exp_mc2}c. This distribution was determined for the first time for almost the entire phase space, except for the region corresponding to low-energy photons correlated with relative angles $\theta_{ij}$ > 170$^\circ$.
This is shown in the experimental spectrum in Fig.~\ref{fig_dalitz_exp_mc2}c which compares with the QED Monte Carlo simulations in Fig.~\ref{fig_dalitz_exp_mc2}a. In most of the possible phase space,  
the green region of Fig.~\ref{fig_dalitz_exp_mc2}c, 
the systematic error is of order 2-3\% while the statistical uncertainty amounts to approximately 3\% for a 15$^{\circ}$ angular binning. The large systematic errors, reaching up to 25\%, are associated with bins in the "corners" of the plot. The systematic uncertainties were estimated by repeating the entire analysis procedure with independent variations of selection cuts. The total systematic uncertainties were then calculated by adding the respective deviations in quadrature. The systematic effect on the sample purity and efficiency is estimated to be about 7\% and 4\%, respectively.





\section{Conclusion and perspectives}

In this work we have presented the first-ever Dalitz distribution for ortho-positronium decay covering nearly the entire accessible phase space. The measurements were carried out using the three-layer J-PET detector, which has demonstrated its capability to register annihilation photons with energies above 30 keV. Although detection threshold limitations exclude very low-energy photons, this study confirms the feasibility to investigate the $o$-$Ps \rightarrow 3\gamma$ decay process with the J-PET detection system and establishes a foundation for future, more comprehensive measurements.



Within our experimental uncertainties, the obtained Dalitz distribution is in agreement with leading-order QED predictions (as shown in Fig.~\ref{fig_dalitz_diffr}).
To go further,
higher statistics and more precise measurements will be necessary to test higher-order QED corrections starting at O($\alpha$)~\cite{Adkins_PhysRep2022, Adkins_PhysRev2005, Adachi2015} and to explore small-scale effects such as the search for any charge-conjugation symmetry breaking with high sensitivity. 
Such advancements are expected to be achievable with the high-acceptance modular J-PET detector~\cite{Moskal_SciAdv2024,Das_BioAlg2025,Faranak_BAMS2024} equipped with silicon photomultipliers, upgraded readout electronics~\cite{Korcyl_Meas2026}, and a dedicated annihilation chamber. 
Performing measurements over the same time range as here with this new  modular detector (assuming 20 times higher sensitivity for the signal reconstruction) equipped with a 5~MBq sodium source, the statistical uncertainty could be reduced by a factor of 10 over the whole available phase space. 
These combined improvements are expected to allow Dalitz plot reconstruction for $o$-$Ps \rightarrow 3\gamma$ decay across a wider kinematic range, 
including low-energy photon regions, and enable probing of higher-order effects.

\section*{Acknowledgements}

We thank Shunsuke Adachi of Kyoto University for help with performing the simulations. We acknowledge support from the National Science Centre of Poland through grants 
MAESTRO no. 2021/42/A/ST2/00423 (P.M.);
OPUS no. 2021/43/B/ST2/02150 (P.M.); 
OPUS24+LAP no. 2022/47/I/NZ7/03112 (E.Ł.S.);
SONATA no. 2023/50/E/ST2/00574 (S.S.);
PRELUDIUM no. 2024/53/N/ST2/04279 (D.K.).
The Ministry of Science and Higher Education through grants no. IAL/SP/596235/2023 (P.M.) and  SPUB/SP/627733/2025 (E.Ł.S.),
European Union within the Horizon Europe Framework Programme through ERC Advanced Grant POSITRONIUM no. 101199807 (P.M.),
The SciMat and qLife Priority Research Areas budget under the program Excellence Initiative – Research University at Jagiellonian University (P.M. and E.Ł.S.).
We also acknowledge Polish high-performance computing infrastructure PLGrid (HPC Center: ACK Cyfronet AGH) for providing computer facilities and support within computational grant no. PLG/2024/017688 and PLG/2025/018762 (M.S.).

\end{document}